\documentclass[prl,11pt,nofootinbib,reprint,showpacs,showkeys]{revtex4-2}
\usepackage{comment}
\usepackage[utf8]{inputenc} 
\usepackage{graphicx,color,overpic,mathtools}
\usepackage{amsthm,amsmath,amssymb,hyperref,mathrsfs}
\usepackage{braket,bm,bbm,setspace}
\PassOptionsToPackage{normalem}{ulem}
\usepackage{ulem} 
\usepackage{soul}
\usepackage{physics}
\usepackage{float}
\usepackage[makeroom]{cancel}
\usepackage[english]{babel}
\usepackage{graphicx}
\graphicspath{ {images/} }
\addto\captionsspanish{}
\hypersetup{
    colorlinks=true,
    pdfborder={0 0 0},
    linkcolor=blue,
    citecolor=blue,
    urlcolor=blue
}

\begin{document}
\vspace*{-2cm}
\hfill 
\vspace*{0.0cm}
\title{Mass inflation without Cauchy horizons}

\author{Ra\'ul Carballo-Rubio}
\affiliation{CP3-Origins, University of Southern Denmark, Campusvej 55, DK-5230 Odense M, Denmark}
\author{Francesco Di Filippo}
\affiliation{Institute of Theoretical Physics, Faculty of Mathematics and Physics, Charles University, V Holešovičkách 2, 180 00 Prague 8, Czech Republic}
\author{Stefano Liberati}
\affiliation{SISSA - International School for Advanced Studies, Via Bonomea 265, 34136 Trieste, Italy}
\affiliation{
IFPU - Institute for Fundamental Physics of the Universe, Via Beirut 2, 34014 Trieste, Italy}
\affiliation{INFN Sezione di Trieste, Via Valerio 2, 34127 Trieste, Italy}
\author{Matt Visser}
\affiliation{School of Mathematics and Statistics, Victoria University of Wellington, PO Box 600, Wellington 6140, New Zealand}

\begin{abstract}
Mass inflation is a well established instability, conventionally associated to Cauchy horizons (which are also inner trapping horizons) of stationary geometries, leading to a divergent exponential buildup of energy. We show here that finite (but often large) exponential buildups of energy are generically present for dynamical geometries endowed with slowly-evolving inner trapping horizons, even in the absence of Cauchy horizons. This provides a more general definition of mass inflation based on quasi-local concepts. We also show that various known results in the literature are recovered in the limit in which the inner trapping horizon asymptotically approaches a Cauchy horizon. Our results imply that black hole geometries with non-extremal inner horizons, including the Kerr geometry in general relativity, and non-extremal regular black holes in theories beyond general relativity, can describe dynamical transients but not the long-lived endpoint of gravitational collapse.
\end{abstract}

\maketitle

\textsl{Introduction.---}The mass inflation instability is an integral part of our understanding of general relativity, playing a crucial role in destabilizing the Cauchy horizons associated with timelike singularities~\cite{Penrose:1968ar,Matzner:1979zz,1982RSPSA.384..301C}, as well as destabilizing the chronological horizons (a sub-class of Cauchy horizons~\cite{Visser:1995cc}) delimiting regions with closed time-like curves. The presence of such an exponential build up of energy is considered a prerequisite to save causality in some general relativity solutions (such as the Kerr black hole), by making singular the boundary of the region violating causality and so, {\em de facto}, by excising the latter from the physical space-time~\cite{Poisson:1990eh,Ori:1991zz,Brady:1995ni}. Hence, mass inflation plays a crucial role in the enforcement of both strong cosmic censorship~\cite{Dafermos:2017dbw,Hollands:2019whz}, and chronology protection. So any possibility of evading mass inflation should be considered with deep suspicion.

Mass inflation is conventionally defined as an infinite divergent exponential buildup of energy~\cite{Poisson:1990eh,Ori:1991zz,Brady:1995ni}. The divergent behavior has been so far associated to stationary geometries, for which inner trapping horizons are always also Cauchy horizons. In all these cases, it is worth noticing that the (finite) exponential buildup of energy and curvature invariants generally leads to a breakdown of the effective description based on general relativity {\em before} any divergence is reached. Indeed, a more physical definition of mass inflation should rely only on the existence of a transient but large exponential buildup --- until high enough curvatures ({\em e.g.}, Planckian) are reached --- regardless of the presence of any mathematical divergence.  This is the novel perspective we adopt in this work.

This shift in perspective is motivated by recent results regarding mass inflation in regular black holes, in which a taming of the inner singularity is postulated on the basis of the idea that quantum gravity (regardless of the specific implementation) should provide a regular description of gravitational collapse~\cite{Bardeen:1968,Hayward:2005gi,Frolov:2014jva,Bonanno:2000ep,Modesto:2005zm,Falls:2010he,Ashtekar:2018lag,Ashtekar:2018cay,Alesci:2019pbs,Platania:2019kyx,BenAchour:2020bdt,BenAchour:2020mgu,BenAchour:2020gon,Bonanno:2023rzk,Benitez:2023wdx} (see also the reviews~\cite{Eichhorn:2022bgu,Platania:2023srt,Ashtekar:2023cod}). It is by now well established that stationary regular black holes with inner horizons also display an initial exponential mass inflation phase --- at least unless the inner horizon is extremal~\cite{Carballo-Rubio:2022kad,Franzin:2022wai} --- that very rapidly brings the regular black hole into a regime where backreaction cannot be neglected~\cite{Brown:2011tv,Carballo-Rubio:2021bpr,Bonanno:2020fgp}, (see also~\cite{Barcelo:2022gii,DiFilippo:2022qkl,Bonanno:2022jjp,Carballo-Rubio:2023kam}).

In this work we show that (large but finite) exponential buildups of energy are also associated with slowly-evolving inner horizons, with the results valid for Cauchy horizons recovered in the stationary limit. This provides a more physical realization of mass inflation implying that even non-eternal black hole spacetimes endowed with a slowly-evolving dynamical inner horizon will generically display a large but finite exponential buildup destabilizing the geometry in short timescales.

Working with two different models of perturbations, the first consisting of two interacting null shells, and the second provided by a null shell interacting with a radiation stream, we will show that both models display the previously mentioned exponential buildup. The first model can be solved analytically, while we shall present numerical results for the second.

\textsl{Adiabatic conditions.---} The ingredients necessary for mass inflation (namely, inner trapping horizons that are perturbed) can be defined also in the presence of rotation, and there is no indication that the latter can prevent the phenomenon~\cite{Bonanno:1995cp,Hamilton:2008zz,Marolf:2011dj}. Hence, for simplicity, we shall work in spherical symmetry, expecting that our results apply also for rotating geometries. An appropriate general parameterization is provided by the line element
\begin{equation}
\text{d}s^2=-e^{-2\Phi(v,r)}F(v,r)\text{d}v^2+2e^{-\Phi(v,r)}\text{d}r\text{d}v+r^2\text{d}\Omega^2,
\end{equation}
where $\text{d}\Omega^2$ is the line element on the unit 2-sphere. We consider black holes with both outer and inner horizons in which the $g_{vv}$ component of the metric vanishes (e.g.,~Reissner-Nordstr{\"o}m or regular black holes). Without significant loss of generality, we can focus on the case with two horizons (regularity at the origin implies an even number of horizons~\cite{Carballo-Rubio:2019fnb}): 
\begin{equation}\label{eq:F_fact}
F(v,r) = e^{\Psi(v,r)} 
\left( 1- \frac{r_{\rm in}(v)}{r}\right)
\left( 1- \frac{r_{\rm out}(v)}{r}\right).
\end{equation}
The two models studied below include an outgoing null thin shell. Outgoing null geodesics satisfy the equation
\begin{equation}
\frac{\text{d}r (v)}{\text{d}v}=\frac{e^{-\Phi(v,r)}F(v,r)}{2},    
\label{eq:out}
\end{equation}
which can be expanded around the position of the inner horizon, $r=r_{\rm in}(v)$, to first order:
\begin{equation}\label{eq:out1st}
\frac{\text{d}r(v) }{\text{d}v}=\frac{e^{-\Phi(v,r_{\rm in}(v))}}{2}\left(\left.\frac{\partial F}{\partial r}\right|_{(v,r_{\rm in}(v))}
\left[r(v)-r_{\rm in}(v)\right]\right)+...
\end{equation}
We define the time-dependent surface gravity of the inner horizon, controlling the \emph{peeling} of null rays around the latter, as~\cite{Cropp:2013zxi}:
\begin{equation}
\kappa_{\rm in}(v)=-\frac{e^{-\Phi(v,r_{\rm in}(v))}}{2}\left.\frac{\partial F}{\partial r}\right|_{(v,r_{\rm in}(v))}=-|\kappa_{\rm in}(v)|.
\end{equation}
This definition reduces to the usual surface gravity of a Killing horizon for stationary geometries~\cite{Cropp:2013zxi}.

\begin{figure}
    \centering
    \includegraphics[width=0.2\textwidth]{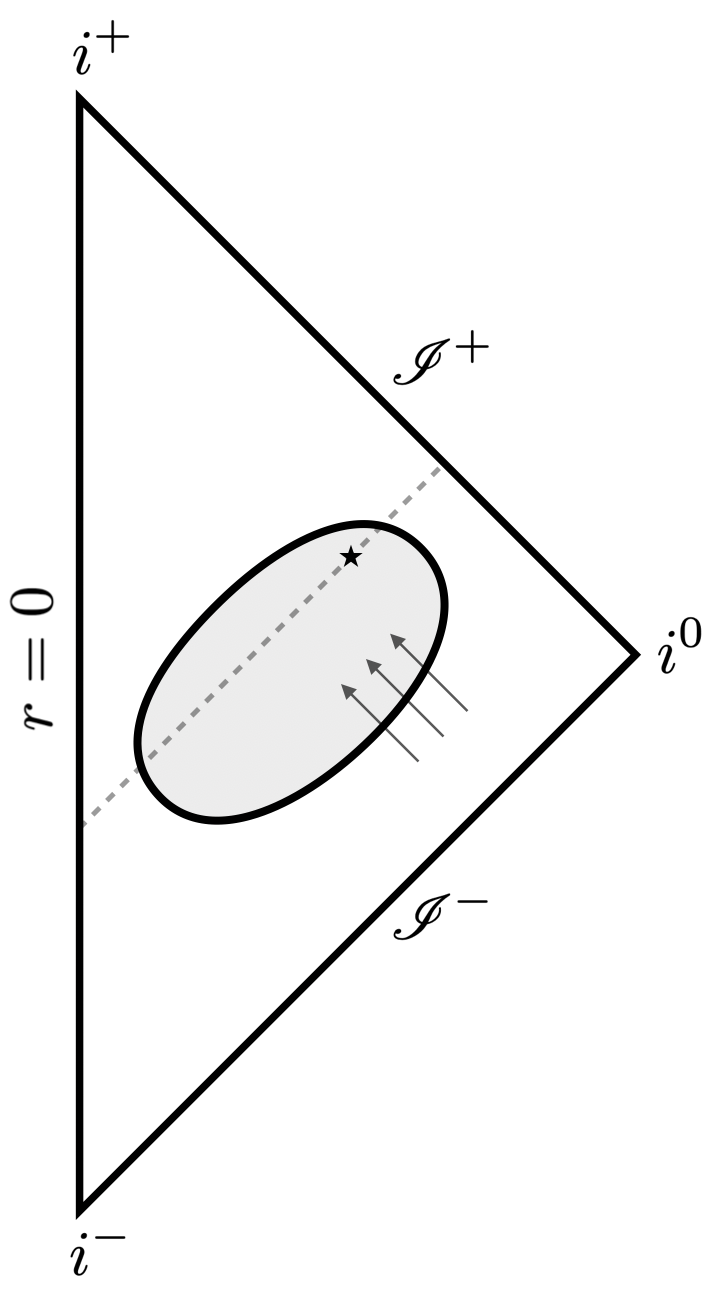}
    \caption{Penrose diagram describing the formation and disappearance of a black hole with both outer and inner horizons, but not Cauchy horizons, marking the boundary of the (topologically closed, as envisaged by Frolov and others~\cite{Frolov:2014jva}) trapped region. The dashed line indicates the placement of an outgoing null thin shell, which approaches the inner horizon exponentially if both adiabatic conditions are satisfied, until a certain critical point marked by the star symbol. The three arrows indicate ingoing perturbations, of which we have considered two different types, either an ingoing null thin shell or a continuous stream of radiation. This perturbation setup is standard in the study of mass inflation.
    }    \label{fig:diag1}
\end{figure}
Eq.~\eqref{eq:out1st} can be recast as a differential equation for the difference $r(v)-r_{\rm in}(v)$ as
\begin{equation}
\frac{\text{d}[r(v)-r_{\rm in}(v)]}{\text{d}v}=
-|\kappa_{\rm in}(v)|\left[r(v)-
r_{\rm in}(v)\right]-\frac{\text{d}r_{\rm in}(v)}{\text{d}v}+...
\end{equation}
If the second term on the right-hand side of the equation above is negligible with respect to the first term, and the surface gravity evolves slowly enough, then the difference $r(v)-r_{\rm in}(v)$ evolves exponentially, exactly as in the stationary situation. These leads to the following adiabatic conditions:\enlargethispage{20pt}
\begin{itemize}
\item \emph{Adiabatic condition for radius of the inner horizon}:
\begin{equation}
\left|\frac{\text{d} r_{\rm in}(v)}{\text{d}v}\right|\ll |\kappa_{\rm in}(v)|
    \left|r(v)-r_{\rm in} (v)\right|.
\label{eq:adbIH}
\end{equation}
\item \emph{Adiabatic condition for surface gravity of the inner horizon}:
\begin{equation}
    \left|\frac{\text{d} \kappa_{\rm in}(v)}{\text{d}v}\right|\ll |\kappa_{\rm in}(v)|^2.
    \label{eq:adbkappa}
\end{equation}
\end{itemize}
These are conditions on the first and second derivatives of $r_{\rm in}(v)$, and thus can be violated or satisfied independently of each other. The first condition requires the specification of an outgoing null geodesic $r(v)$ for its evaluation. The second condition is the analogue of the condition discussed for the outer horizon in~\cite{Barcelo:2010pj,Barcelo:2010xk}. Under these two conditions, we can write
\begin{equation}\label{eq:expsol}
 r(v)\approx r_{\rm in}(v)+\left[r(v_0)-r_{\rm in}(v_0)\right]e^{-|\kappa_{\rm in}(v)|(v-v_0)},
\end{equation}
with the initial condition $r(v_0)\in(r_{\rm in}(v_0),r_{\rm out}(v_0))$. Inserting Eq.~\eqref{eq:expsol} in the adiabatic condition for the radius of the inner horizon, we obtain
\begin{equation}\label{eq:1stad_v}
   \left | \frac{\text{d} r_{\rm in}{(v)}}{\text{d}v}\right | \ll |\kappa_{\rm in}(v)|\left[r(v_0)-r_{\rm in}(v_0)\right]e^{-|\kappa_{\rm in}(v)|(v-v_0)}.
\end{equation}
This equation illustrates that, even if the adiabatic condition is always satisfied, the condition for slow variation of the inner horizon will always eventually cease to be valid as long as $\text{d}r_{\rm in}/\text{d}v$ is non-zero. This will happen at some time $v_\star$ that, given the assumed adiabatic evolution of the surface gravity, is approximately given by the explicit formula
\begin{equation}
\label{E:v-star}
v_\star \approx v_0 + \frac{1}{|\kappa_{\rm in}(v_\star)|} 
\ln \left\{\frac{
|\kappa_{\rm in}(v_\star)|\; \left[r(v_0)-r_{\rm in}(v_0)\right]} {\left|\text{d} r_{\rm in}/\text{d}v\right|_{v_\star}} 
\right\}.
\end{equation}
This observation is important for our discussion below.

\textsl{Analytical results.---} Aside from the outgoing null thin shell, let us introduce an ingoing null thin shell, which is a standard setup to discuss mass inflation~\cite{Dray:1985yt,1990CQGra...7L.273B}. The interaction between outgoing and ingoing null thin shells leads to exponential mass inflation in the region of spacetime in between the two shells. Despite the original analysis being focused on general relativity, some of the results can be obtained on a purely geometrical grounds without the need of specifying the dynamics of the theory \cite{Carballo-Rubio:2018pmi,DiFilippo:2022qkl}. In particular, independently of the theory, the mass $m_{\rm f}(v_\times,r_\times)$ in between the ingoing and outgoing null shells after the crossing of the two shells at $r_\times=r(v_\times)$ is given by~\cite{Carballo-Rubio:2018pmi,DiFilippo:2022qkl}
\begin{align}\label{eq:dtr_mass}
m_{\rm f}(v_\times,r_\times) &= m_{\rm i}(v_\times,r_\times) + m_{\rm in} (v_\times,r_\times) +m_{\rm out}(v_\times,r_\times)\nonumber\\ 
&- \frac{2m_{\rm in} (v_\times,r_\times) m_{\rm out}(v_\times,r_\times)}{r_\times F_{\rm i}(v_\times,r_\times)},
\end{align}
where $m_{\rm i}(v_0,r_0)$ is the mass in between the two shells prior to the crossing at $r=r_\times$ while $F_{\rm i}(v_\times,r_\times)$ is the metric function in between the two shells prior to the crossing which, using Eq.~\eqref{eq:F_fact}, we can write in terms of $\left(v_\times,r(v_\times)\right)$ as
\begin{equation}
F^\times_{\rm i}= \frac{e^{\Psi(v_\times,r(v_\times))}}{r_{\rm in}(v_\times)} 
\left( 1- \frac{r_{\rm out}(v_\times)}{r_{\rm in}(v_\times)}\right)
\left( r(v_\times)- r_{\rm in}(v_\times)\right).
\end{equation}

On the other hand, $m_{\rm in} (v_\times,r_\times)$ and $m_{\rm out}(v_\times,r_\times)$ measure the jump of the mass function across the ingoing and the outgoing shell. The specific properties of the theory under consideration only enter in the determination of $m_{\rm in} (v_\times,r_\times)$ and $m_{\rm out}(v_\times,r_\times)$. In the following, we will simply assume that these quantities are proportional to the energy of the shells.
The crossing time $v=v_\times$ can be chosen so that the quantity $r(v_\times)- r_{\rm in}(v_\times)$ can be as small as possible. Hence, the mass $m_{\rm f}(v_\times,r_\times)$ in Eq.~\eqref{eq:dtr_mass} will typically grow large for generic perturbations. 

Nonetheless, we saw that the exponential behavior characteristic of static situations requires that \emph{both} of the aforementioned adiabatic conditions, Eqs.~\eqref{eq:adbIH}-\eqref{eq:adbkappa} are satisfied. However, the first adiabatic condition for the evolution of the inner horizon cannot be maintained indefinitely, not least for the backreacton of the aforementioned exponential mass inflation. This implies that any exponential behavior is regulated by the variation in the position of the inner horizon, as the minimum value of the function $F(v_\times,r(v_\times))$ that can be reached during the period of exponential growth, $F^\star_{\rm i}=F_{\rm i}\left(v_\star,r(v_\star)\right)$, is given by
\begin{equation}\label{eq:F_fact_vstar}
F^\star_{\rm i} = \frac{e^{\Psi(v_\star,r_{\rm in}(v_\star))}}{r_{\rm in}(v_\star)} 
\left( 1- \frac{r_{\rm out}(v_\star)}{r_{\rm in}(v_\star)}\right)
\frac{\left|\text{d}r_{\rm in}/\text{d}v\right|_{v_\star}}{|\kappa_{\rm in}(v_\star)|},
\end{equation}
where $v=v_\star$ indicates the time approximately given in Eq.~\eqref{E:v-star}.

Using Eqs.~\eqref{eq:dtr_mass} and~\eqref{eq:F_fact_vstar}, the mass $m_{\rm f}(v_\star,r_\star)$ increases exponentially up to a maximum value
\begin{equation}
M_{\rm max}\approx \frac{r_{\rm in}(v_\star)\;|\kappa_{\rm in}(v_\star)|}{
\left|\left.\text{d}r_{\rm in}(v)/\text{d}v\right|_{v_\star}\right|} \; 
\frac{2 m_{\rm in} (v_\times, r_\times) m_{\rm out}(v_\times, r_\times)}{r_\times}.
\label{eq:regM}
\end{equation}
Note that this can be factorized into the form $M_{\rm max}\simeq f_1(v_{\star}) \; f_2(v_\times,r_\times)$, with one function depending on the end of exponential mass inflation, and the other only on the null-shells crossing.

Eq.~\eqref{eq:regM} can be understood as the regularized version of $M_{\rm max}=\infty$ that is obtained in the static case, where the regulator comes from the exponential approximation ceasing to be valid due the non-zero  value of $\left.\text{d}r_{\rm in}(v)/\text{d}v\right|_{v_\star}$. As a consistency check, for $\left.\text{d}r_{\rm in}(v)/\text{d}v\right|_{v_\star}\to 0$ we recover the result $M_{\rm max}=\infty$.

We will now compare these analytical results with numerical results based on an extension of the Ori model~\cite{Ori1991}.

\textsl{Numerical results.---} Let us introduce a slightly different perturbation type wherein we maintain an outgoing null shell, but replace the ingoing null shell with a continuous stream of energy. This configuration was originally examined by Ori~\cite{Ori1991} to investigate the instability of Reissner-Nordström black holes, and has been applied later to more general situations~\cite{Brown:2011tv,Carballo-Rubio:2021bpr,Bonanno:2020fgp}.

The gluing conditions for the Ori model along the outgoing shell can be written for any asymptotically flat geometry. These conditions lead to differential equations for the Misner-Sharp mass $M(v)=-1+rF(v,r)/2$ inside the outgoing shell that can be solved numerically. The pressureless nature of the shell leads to the continuity of $T_{\mu\nu}  s^\mu s^\nu$, where $s^\mu=\left( {2}/{F},1,0,0 \right)$ is an outgoing null vector field normal to the shell~\cite{Barrabes:1991ng}. This continuity condition implies~\cite{Carballo-Rubio:2021bpr}
\begin{equation}\label{eq:junction}
\left. \frac{1}{F_+}\frac{\partial M_+}{\partial v}\right|_{r=R(v)}=\left.\frac{1}{F_-}\frac{\partial M_-}{\partial v}\right|_{r=R(v)}\,,
\end{equation}
where $R(v)$ denotes the radius of the outgoing shell and the $+$ ($-$) index indicates that the corresponding quantity must be evaluated inside (outside) the outgoing null thin shell (see Fig.~\ref{fig:diag1}).

Our results do not depend on the specific regular black hole considered, but only on the existence of an inner horizon, which is a generic feature associated with the requirement of regularity~\cite{Carballo-Rubio:2019fnb}. For simplicity, let us consider the Bardeen metric~\cite{Bardeen:1968},
\begin{equation}
F_\pm(v,r)=1-\frac{2r^2m_\pm(v)}{\left(r^2+\ell^2\right)^{3/2}}.
\end{equation}
This metric has two horizons, a single extremal horizon or no horizons, depending on the value of $m_\pm(v)$~\cite{Bardeen:1968}.  

The functional form of $m_-(v)$ is loosely constrained to describe the situation depicted in Fig.~\ref{fig:diag1}, namely a geometry with no horizons at early and late times, and an ingoing stream of radiation. Without loss of generality, let us consider
\begin{align}\label{eq:din_Price}
    m_-(v)=&\left[M_\infty+\delta m(v)\right]\nonumber\\
    &\times \frac{\tanh{\left[ s_1(v-v_i) \right]} -\tanh{\left[ s_2(v-v_f) \right]}}{2} ,
\end{align}
where $\delta m(v)$ is a perturbation of the mass, $v_i\ll v_f$, and $s_1$ and $s_2$ positive constants. Eq.~\eqref{eq:din_Price} describes an approximately flat spacetime for $v\ll v_i$ and $v\gg v_f$, while for $v_i\ll v\ll v_f$ it describes a black hole accreting mass according to $\delta m(v)$. We are using hyperbolic tangents as interpolating functions to provide a specific realization, but other functions (possibly of compact support) can be considered.

\begin{widetext}

\begin{figure}[!htbp]
\begin{minipage}{0.49\textwidth}

    \centering
    \includegraphics[width=0.9\linewidth]{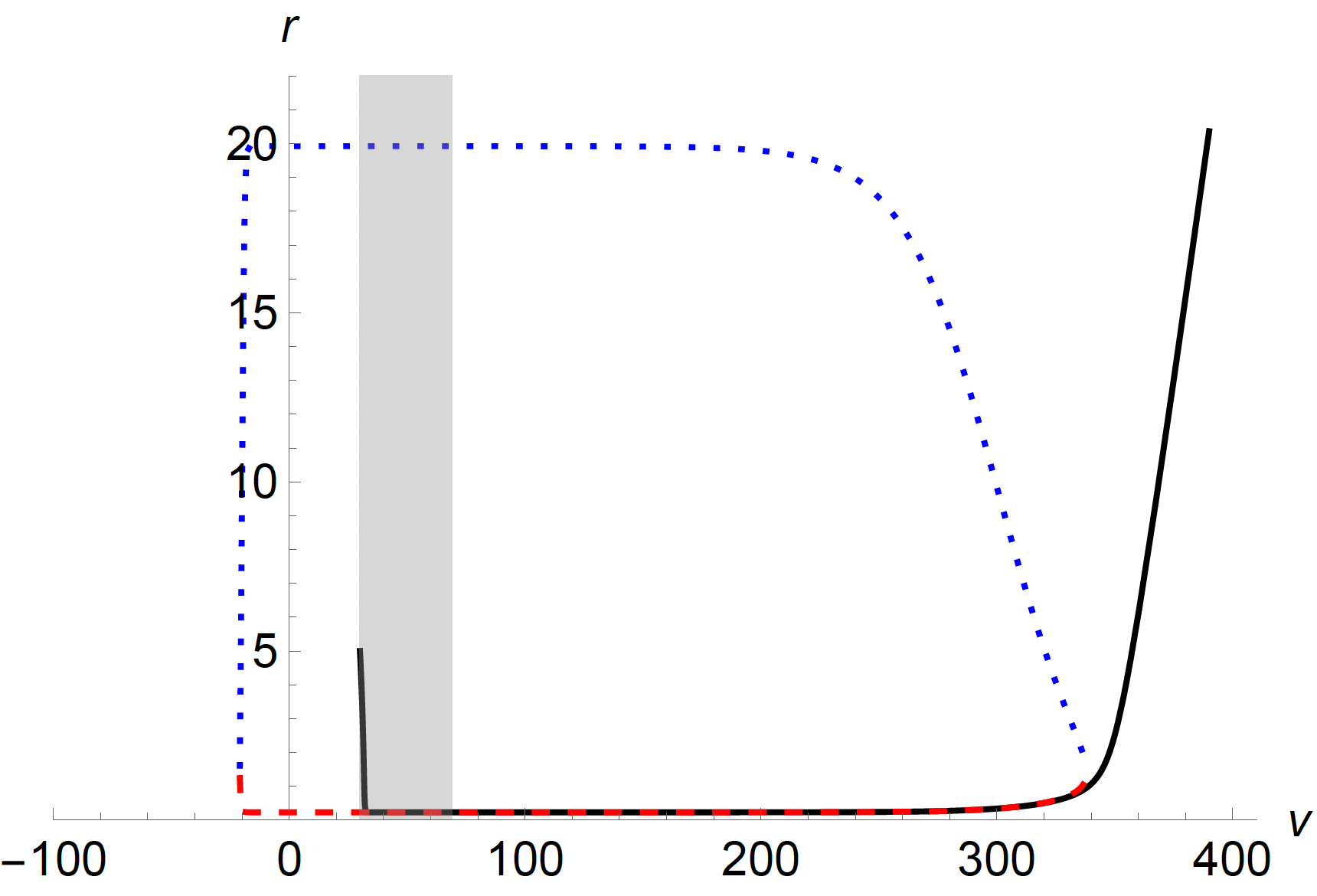}

\end{minipage}
\begin{minipage}{0.49\textwidth}

\centering\includegraphics[width=.9\linewidth]{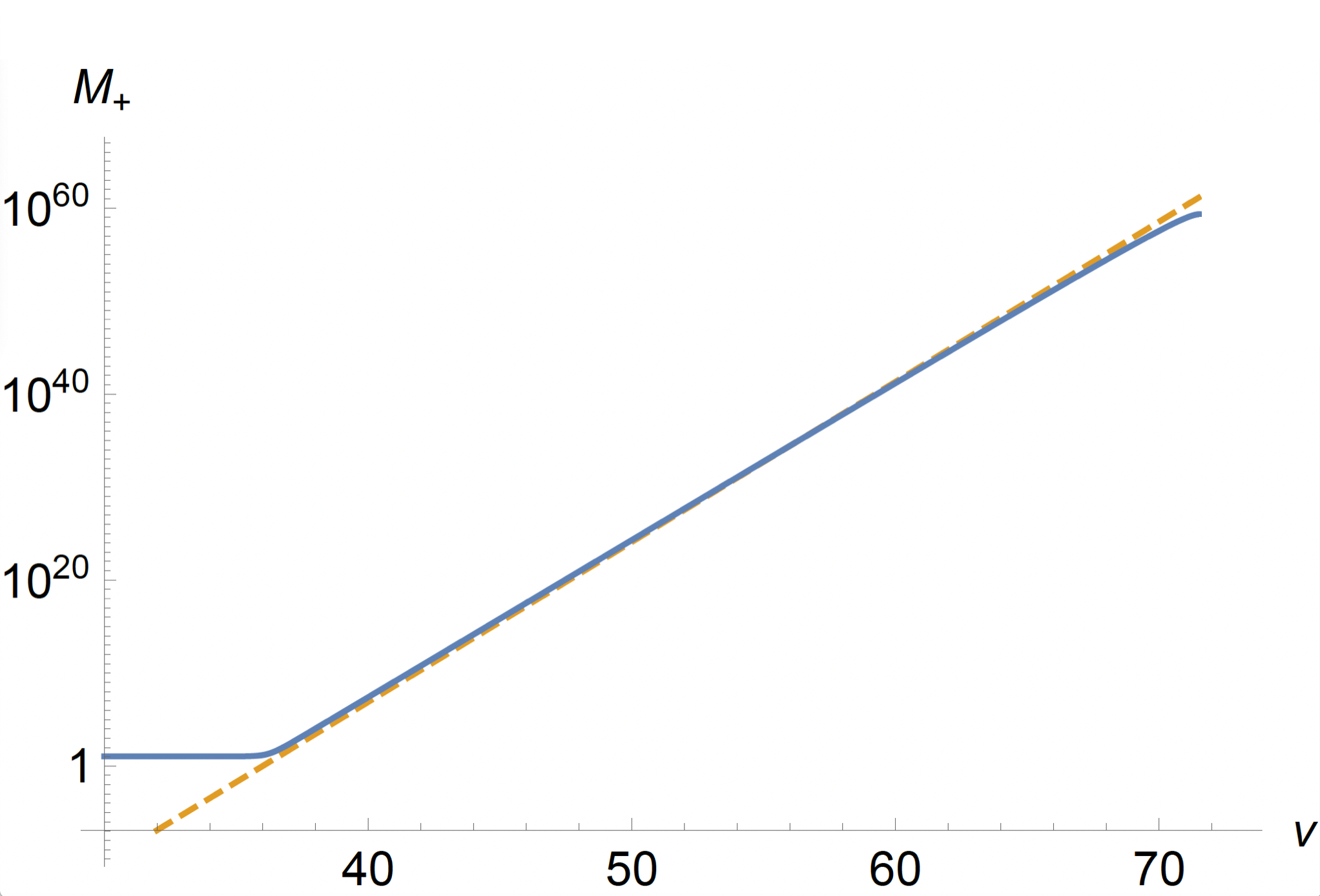}

\end{minipage}

    \caption{\emph{Left:} Specific realization of Fig.~\ref{fig:diag1} for the Misner-Sharp mass in Eq.~\eqref{eq:din_Price} and parameters $M_\infty=10,\,\ell=1,\,\beta=1,\,\gamma=2,\,v_i=-20,\,v_f=300,\,s_1=1,\,s_2=1/40 $. Outer and inner horizons are indicated by the blue dotted line and the red dashed line, respectively, and the outgoing null thin shell by the solid black line. The radius of the outgoing shell is set initially at $R(v=30)=5$. The shaded region marks the interval of time for which $M_+(v)$ is plotted in the right-hand panel. \emph{Right:} Misner-Sharp mass $M_+(v)$ in the region interior to the outgoing shell for the the initial condition $m_-(v=30)=M_\infty+1=11$. The dashed line corresponds to the exponential mass inflation for a stationary regular black hole with the same parameters. The exponential buildup of $M_+(v)$ pushes the linear approximation beyond its regime of validity in a short interval of time, as in the stationary case.
}
    \label{fig:Radius}
    
\end{figure}
\end{widetext}

\enlargethispage{50pt}
The mass in the interior region, $M_+(v)$, can be obtained integrating Eq.~\eqref{eq:junction} numerically. We perform this integration for a finite interval of time contained within the trapped region in which the adiabatic conditions are satisfied, as shown in Fig.~\ref{fig:Radius}. For concreteness, let us assume that the usual Price law~\cite{Price1972,Price:1972pw,Price:2004mm},
\begin{equation}\label{eq:etern_Price}
    \delta m(v)=-\frac{\beta}{v^\gamma},
\end{equation}
 is satisfied in this interval, which lead to the numerical evolution of $M_+(v)$ shown in Fig.~\ref{fig:Radius}. Actually, the exponential in Fig.~\ref{fig:Radius} is generic for any perturbation $\delta m(v)$ that decays slower than $e^{-|\kappa_{\rm in}| v}$, as in the stationary case~\cite{Carballo-Rubio:2021bpr}. 

As shown in these figures, the dynamical evolution of the mass follows the same exponential instability of the stationary case whenever the adiabatic conditions are satisfied, thus matching the analytical results discussed previously.

\textsl{Conclusions.---} We have shown that the exponential buildup characteristic of the mass inflation instability is not limited to stationary black hole spacetimes, but extends to dynamical spacetimes, as long as the inner horizon is non-extremal and the geometry is evolving sufficiently slowly as encapsulated in two adiabatic conditions which become exact in the stationary limit, so allowing us to recover the standard results for Cauchy horizons. 

While the mass inflation backreaction eventually imples a breakdown of one or both adiabatic conditions, this  generically happens after mass inflation has triggered a rapidly evolving phase resulting into the perturbative treatment ceasing to be valid.

The analysis presented herein is restricted to spherical symmetry, nonetheless the ingredients leading to mass inflation in our setting are present, and known to generically lead to the same phenomenon, also in the presence of rotation. Hence, there is no reason to expect that inner horizons in rotating black holes would behave any differently.

It has been conjectured that mass inflation might lead to a null-like singularity in the black hole interior without affecting the exterior geometry~\cite{Dafermos:2012np}. Such a conjecture is far from being demonstrated (the mass inflation instability and its large backreaction would still be present even close to a null singularity) and assumes no resolution of the singularity by quantum gravitational effects. The latter could lead to different scenarios~\cite{Carballo-Rubio:2019fnb,Carballo-Rubio:2019nel}, including regular black holes with inner-extremal cores~\cite{Carballo-Rubio:2022kad,Franzin:2022wai}, Simpson-Visser cores (also called hidden wormholes)~\cite{Simpson:2018tsi,Mazza:2021rgq}, or bouncing cores~\cite{Barcelo:2020mjw,Barcelo:2022gii} that may result into a horizonless ultra-compact object of the same family of the initial regular black hole~\cite{Carballo-Rubio:2022nuj}.  

The implications are striking: generic black holes with (non-extremal) inner horizons will always keep evolving in a timescale controlled by $1/\kappa_{\rm in}$, and cannot be the endpoint of a stellar collapse. It is generally believed that astrophysical black holes are well described by a quasi-stationary Kerr metric, possibly with a regularized Planckian core. Our results challenge this expectation and show that determining the endpoint of stellar collapse is an inevitable open question.

\begin{acknowledgments}
\textsl{Acknowledgments.--} RCR acknowledges financial support through a research grant (29405) from VILLUM fonden.
FDF acknowledges financial support by Primus grant PRIMUS/23/SCI/005 from Charles University and by GAČR 23-07457S grant of the Czech Science Foundation. MV was supported by the Marsden Fund, via a grant administered by the Royal Society of New Zealand. 
\end{acknowledgments}

\bibliographystyle{unsrt}
\bibliography{refs}

\end{document}